# Structure and Quantum Well States in Silicene Nanoribbons on Ag(110)

Baojie Feng[1], Hui Li[1], Sheng Meng[1,2], Lan Chen[1]*, and Kehui Wu[1,2]*

*1 Institute of Physics, Chinese Academy of Sciences, Beijing 100190, China*

*2 Collaborative Innovation Center of Quantum Matter, Beijing 100871, China*

Email: lchen@iphy.ac,cn (L. Chen) & khwu@iphy.ac.cn (K. Wu)

**We have performed scanning tunneling microscopy/spectroscopy (STM/STS) study and first-principles calculations to investigate the atomic structure and electronic properties of silicon nanoribbons (SiNRs) grown on Ag(110). Despite of the extensive research on SiNRs in the last decades, its atomic structure is still not fully understood so far. In this report we determine that the structure of SiNRs/Ag(110) is armchair silicene nanoribbon with reconstructed edges. Meanwhile, pronounced quantum well states (QWS) in SiNRs were observed and their energy spectrum was systematically measured. The QWS are due to the confinement of quasiparticles perpendicular to the nanoribbon and can be well explained by the theory of one-dimensional (1D) "particle-in-a-box" model in quantum mechanics.**

A two dimensional silicon sheet with honeycomb structure, namely silicene, has attracted much attention recently [1–3]. Compared with graphene, silicene has a larger spin-orbit coupling strength, which may lead to detectable quantum spin Hall effect (QSHE). [3] In addition, the buckled structure of silicene results in a pronounced response to the external field, leading to more controllable electronic and magnetic properties which will benefit its further device applications. [4] Silicene has been successfully grown on different substrates including Ag(111), $ZrB_2$, and Ir(111). [5–8] The novel physical properties of silicene, such as chiral Dirac fermions, [9,10] structural phase transition [11] and intriguing flat band below the Fermi level [12] have been discovered experimentally.

In the case of graphene, patterning graphene sheets into nanoribbons has been shown effective to create an energy bandgap, owing to the quantum confinement within the finite ribbon width [13-15]. One can expect similar effect if silicene is patterned into nanoribbons. Indeed, even before the experimental realization of a silicene sheet, Cahangirov *et al.* has theoretically proposed that silicene nanoribbon may possess interesting properties such as energy gap and magnetic ordering. [2] Interestingly, in contrast to graphene where nanoribbons are difficult to obtain, one-dimensional silicene nanoribbons (SiNRs) have been found to spontaneously form on Ag(110) [16-22] and Au(110) [23] surfaces. It was also reported in an angle resolved photoemission spectroscopy (ARPES) study that SiNRs on Ag(110) exhibit a Dirac cone at the $\bar{X}$ point of the Brillouin zone [24], suggesting that they could possibly be graphene-like silicene nanoribbons. However, to date, the atomic structure of the SiNRs/Ag(110) is still under debate. Several structural models have been theoretically proposed for these SiNRs, such as zigzag silicene nanoribbon [25, 26], armchair silicene nanoribbon, [27] and so on [28]. Among them, the zigzag model was considered to be accorded with the APRES results. But none of them can fully agree with the experimental scanning tunneling microscopy (STM) images. On the other hand, the electronic properties of SiNRs have been little explored apart from the above mentioned ARPES study.

Here, we report on a study on the atomic structure of SiNRs/Ag(110) by low temperature STM/STS experiments combined with first-principles calculations. Our results suggest that the SiNRs are armchair silicene nanoribbon with reconstructed edges. The simulated STM images can match our atomically resolved STM images perfectly. Moreover, we observed pronounced quantum well states (QWS) in the SiNRs. The energy spectrum of the QWS was systematically measured and explained based on the 1D "particle-in-a-box" model in quantum mechanics. These findings are helpful to understand the structure and properties of SiNRs, which is important for the future nanoelectronic application based on silicene nanoribbon.

Experiments were carried out in a home-built low temperature STM-MBE system with base pressure of $5\times10^{-11}$ Torr. Single crystal Ag(110) was cleaned by Ar$^+$ ion sputtering and annealing cycles. Silicon was evaporated from a heated wafer ($\approx$1200 K) onto the clean Ag(110) substrate. The flux of silicon was kept at 0.08-0.1 ML/min. Here one monolayer is defined as the atomic density of an ideal silicene sheet, *i.e.* $1.69\times10^{15}$ cm$^{-2}$. [7] The STS data were acquired using a lock-in amplifier by applying a small sinusoidal modulation (20 mV and 677 Hz) to the bias voltage. All the STM data presented in this letter were acquired at 77 K. The bias voltage is applied to the tip with respect to the sample.

The SiNRs can form on Ag(110) in a wide temperature range, from room temperature to approximately 500 K. There are mainly two kinds of SiNRs with width of 1.0 nm and 2.0 nm, respectively. [17] When submonolayer silicon atoms, for example 0.5 ML, are deposited onto Ag(110) at room temperature, SiNRs with width of 1.0 nm are obtained, as shown in Fig.1(a) and (b). As the substrate temperature increases, the 2.0 nm wide SiNRs appear and their lengths increase as well. When the substrate temperature reaches 440 K, Ag(110) surface is dominated by 2.0 nm wide SiNRs whose length is typically hundreds of nanometers, as shown in Fig.1(c) and (d). From the high resolution STM images (Fig.1(b), (d) and (e)), we find that the 1.0 nm and 2.0 nm wide SiNRs consist of two and four rows of protrusions, respectively, which are closely packed along the [$1\bar{1}0$] direction of Ag(110). These results are consistent with previous reports. [17]

Up to now, there have been several theoretical proposals for the structure model of 2.0 nm SiNRs/Ag(110), but none of them fit with the high-resolution STM images. For example, A. Kara *et al.* proposed a zigzag model for the 2.0 nm wide SiNRs. [25, 26] However, their simulated LDOS in real space showed a rectangular symmetry, and the period along the nanoribbon is one lattice constant of silicene-1×1, *i.e.* 0.38 nm. These features are obviously in contradiction with our high resolution STM image, as shown in Fig. 1e. In this image one can see that the protrusions along the two edges of the ribbon are not mirror-symmetric, but shifted half of a period along the ribbon direction with respect to each other. In addition, in this high-resolution image we can see atomic resolution of the Ag(110) substrate as well. By comparing the lattice of the Ag(110) substrate and the SiNR we found that the periodicity along the nanoribbon is exactly $2a_{Ag}$, where $a_{Ag}$ is the lattice constant of Ag(110) along the [1$\bar{1}$0] direction. Another structural model proposed by C. Lian *et al.* is the armchair silicene nanoribbon. [27] This model produces a period of 5.87 Å along the SiNRs, which agrees with the experimental value. However, the LDOS simulations of this model also showed a rectangular symmetry, in contrast to the experiments. For the structure of 1.0 nm SiNRs/Ag(110), the detailed theoretical prediction is still lacking.

We note that recently, there are emerging debates on the growth of Si on Ag. Several papers reported that the morphology of the Ag(110) substrate is substantially modified during the growth of Si [29,30], suggesting possible alloying of Si with Ag. Consequently, although our theoretical model can explain our STM image well (as discussed below), we have no solid evidence to exclude the possibility of alloying in this system. However, in our experiment, we have found that all SiNRs disappeared after annealing to 700 K, which indicates that the Si and Ag atoms are unlikely to form alloy on the surface. Suppose that the SiNRs are Si-Ag alloy, annealing at high temperature is unable to separate the two types of atoms. The growth behavior of SiNRs on Ag(110) is analogous to silicene 3×3 reconstruction grown on Ag(111) where the Ag(111) substrate has also been modified during the growth of Si [31]. However, both theoretical calculations [5,32] and extensive experiments [33,34] have proven the validity of the structural model of silicene 3×3 on Ag(111). Therefore we suggest that the most possible structure of SiNRs on Ag(110) consists of only Si atoms.

To determine the atomic structure of SiNRs, we performed first-principles calculations on both 1.0 nm and 2.0 nm SiNRs, respectively. A supercell of 6×2 Ag(110) surface with 6 layers of Ag atoms is used to mimic the substrate in the computational box, which is covered by the 1.0 nm or 2.0 nm SiNRs, and the vacuum distance is set as 20 Å. The Perdew-Burke-Ernzerhof (PBE) [35] exchange-correlation functional was employed, as well as the projector augmented wave (PAW) pseudopotentials combined with plane wave basis sets with energy cutoff of 250 eV. For geometry optimization, the surface Brillouin zone was sampled by 2×8×1 k-points using the Monkhorst-Pack scheme, and the optimized structure was relaxed until the maximum force on each atom is less than 0.01eV/Å. For the calculations of electronic properties of SiNR/Ag model, 2×16×1 k-points were chosen. All the calculations were carried out using the Vienna *ab initio* Simulation Package. [36]

Starting from more than 50 initial adsorption geometries with different widths of SiNRs, two possible structures were found most close to the STM images, as shown in Fig.2(a)-(d). For both types of structures, the center part of the nanoribbon is perfect honeycomb structure, along the [1$\bar{1}$0] direction of Ag, with armchair edges. Unlike the in-commensurate structure of √3×√3 monolayer silicene on Ag(111) substrate [7,11], silicon nanoribbons form commensurate structure on Ag(110), with the half of the Si atoms on the bridge sites between two adjacent silver rows. As we know, a free-standing silicene has a smaller lattice constant (3.86 Å) than the column-column distance (4.08 Å) on the Ag(110) surface. The tensile stress caused by the mismatch between the lattices of silicene and Ag substrate increases with the increasing width of the SiNRs along Ag [001] direction. As a result, the width of silicene ribbon is limited to narrower than 2.0 nm on Ag(110) surface. Moreover, unlike previously proposed models of SiNRs, in our model the edges of 2.0 nm SiNRs are reconstructed, [Fig. 2(c) and (d)], resulting in a highly buckled structure similar to the √3×√3 structure of monolayer silicene. [11] This is probably due to the tendency for Si atoms to form sp$^3$ hybridization instead of sp$^2$ hybridization. For the 1.0 nm SiNR, one side of reconstructed edge is the same as that in the 2.0 nm wide SiNR, and the other side is a distorted armchair structure. The Si atoms buckled upwards (red atoms in Fig. 2(a)-(d)) can be probed by STM as protrusions in the topographic images. For the 1 nm and 2 nm SiNRs models, the lateral distance between the highly buckled Si rows at the two edges are 1.5 and 0.6 nm, as indicated by the black arrows in Fig.2(a) and (c). These values are in perfect agreement with the distances measured in STM images. The simulated STM images correspond well with the experimental ones, as shown in Fig.2(e) to (l), which also supports our structure models.

In order to reveal the electronic properties of SiNRs we further performed STS measurements of the local density of states (LDOS) of the SiNRs. To avoid influence from neighboring nanoribbons, we select an isolated SiNR with 2.0 nm width. The dI/dV maps of bias voltage from -0.5 V to -4.0 V with interval of 0.1 V were obtained to show the distribution of LDOS in real space. Fig. 3(a)-(d) are four typical examples. At low bias, the LDOS is mainly distributed around the two edges of the SiNR, with the center being depressed (Fig.3(a)). With increasing bias voltage, the LDOS at the edges disappear. Meanwhile, one, two and three bright strings appear at the center of the nanoribbon successively (Fig.3(b), (c) and (d)). To show the evolution of LDOS as a function of bias voltage, we plotted the line profiles across the nanoribbon (along the black line in Fig.3(a)) for images measured from -0.5 V to -4.0 V, which are shown in Fig.3(e). One can unambiguously observe the evolution of peaks from one, two to three with the bias voltage increasing, as indicated by the black dotted lines in Fig.3(e).

The oscillating patterns inside the SiNRs can be assigned to quantum well states (QWS), as explained by the one dimensional "particle-in-a-box" model. Wave functions in a 1D quantum well are sinusoids with wavenumber $k_n=n\pi/L$, where $n$ is the quantum number and $L$ is the width of the quantum well. However, we found most line profiles are not uniform, and the waves are even asymmetric with respect to the center. They should be attributed to the overlap of quantum well states [37] and the edge states, *i.e.*, $dI/dV(V) \propto \sum_n c_n(V)|\varphi_n(k)|^2 + |\psi_{edge}|^2$. Here, $c_n(V)$ are the coefficients for each eigenstate $\varphi_n(k)$ of the QWS while $\psi_{edge}$ is the wave function of the edge state of SiNRs. From Fig.3(e), we find that the oscillations of the LDOS at energies -1.6 V, -2.5 V and -3.8 V are more regular and symmetric. These states are considered as the pure eigenstates without overlaps with neighboring states. According to basic theory of quantum mechanics, the eigenenergy of the 1D infinite quantum well is $E_0 + n^2\pi^2\hbar^2/2m^*L^2 = E_0 + \hbar^2 k^2/2m^*$, where $E_0$ is the onset energy and $m^*$ is the effective mass of electrons. That is to say, the energy-momentum dispersion is parabolic. In the fitting process, we set $L$ as the apparent width of the nanoribbon, *i.e.*, $L$=2.0 nm. A parabolic fit to the dispersion relation, shown in Fig.3(f), yields the onset energy $E_0$ of 1.36±0.04 eV and effective mass $m^*$ of (0.34±0.01)$m_e$. The effective mass is very close to those reported in 1D metal chains and gratings [37–39], indicating that the QWS might originate from metallic states. We also calculated the density of states (DOS) of the relaxed SiNRs without Ag(110) substrate, as shown in Fig.2(m) and (n). The total DOS of SiNRs mainly come from the *p* orbital of silicon, and show a metallic character from -5 eV to 5 eV, in good accordance to our experiments on observation of QWS.

It should be noted that in a previous work, F. Ronci *et al.* have already observed the n=1 and n=2 eigenstates of the QWS in SiNRs using STS [22]. However, because only the lowest two eigenstates have been observed, they cannot exclude the possibility of edge states in SiNRs and they are unable to extract the energy-momentum dispersion of the QWS. In our experiments, we have observed the n=3 eigenstate of the QWS and successfully fitted the parabolic energy-momentum dispersion. Thus our results unambiguously prove the existence of QWS in SiNRs.

In previous STS experiments on pure Ag(110), standing waves stemming from the highest unoccupied surface states $S_2$ [40] and surface-projected bulk band [41] had been reported. In our experiment, we have observed the $S_2$-derived standing wave patterns on bare Ag(110) and the QWS on SiNRs simultaneously, as shown in Fig.3(c) and (d). Their wavelengths and energy ranges are obviously different. The surface state $S_2$ is located at 1.7 eV above the Fermi level while the first eigenstate of QWS appear at 0.8 eV above the Fermi level. More importantly, the wavelengths of $S_2$-derived standing wave patterns increase with the bias voltage, which is in contrast to the case of QWS in SiNRs. So we can exclude the possibility that the QWS on SiNRs originate from the surface states $S_2$ of Ag(110). In our experiment, we did not observe the standing wave patterns originating from the bulk bands on bare Ag(110). This may be attributed to the relatively high temperature (77 K) in our STS experiments and the interference patterns are probably smeared out. On the other hand, on bare surface areas of Ag(110) confined between two SiNRs with 2.0 nm separation, we have never observed similar QWS as those on SiNRs (data not shown here). Therefore, the possibility that the QWS originate from the surface-projected bulk bands can also be ruled out. As a result, we conclude that our observed QWS are an intrinsic character of SiNRs.

At last, we emphasize that our structural model is already a reconstructed silicene model. The central part of the ribbon is honeycomb silicene structure, but there are two rows of Si adatoms at the edges (We failed to construct a model with pure 2D silicene ribbon structure). Since the ribbon is narrow, the reconstructed edge should have pronounced influence on the electronic structure, and therefore most likely the Dirac cone structure for pristine silicene would no longer exist. However, the metallic nature of SiNRs is still interesting which makes SiNRs a promising material in future nanoscale Si devices.

In summary, we have investigated the structure and electronic properties of the SiNRs epitaxially grown on Ag(110) using STM/STS. Combined with first-principles calculations, the structure of SiNRs/Ag(110) has been determined as armchair silicene nanoribbon with reconstructed edges. Due to the confinement of quasiparticles perpendicular to the nanoribbon, pronounced QWS can be observed in these SiNRs, which can be explained by a simple 1D "particle-in-a-box" model in quantum mechanics. Detailed analysis have unambiguously shown that the QWS originate from the metallic states of SiNRs, instead of bands of Ag(110).

**Acknowledgement:**

This work was supported by the MOST of China (Grants No. 2012CB921703, 2013CB921702,

2013CBA01600), and the NSF of China (Grants No. 11334011, 11174344, 11322431, 91121003).

**Author contributions:**

**Additional information**

**Competing financial interests:** The authors declare no competing financial interests.

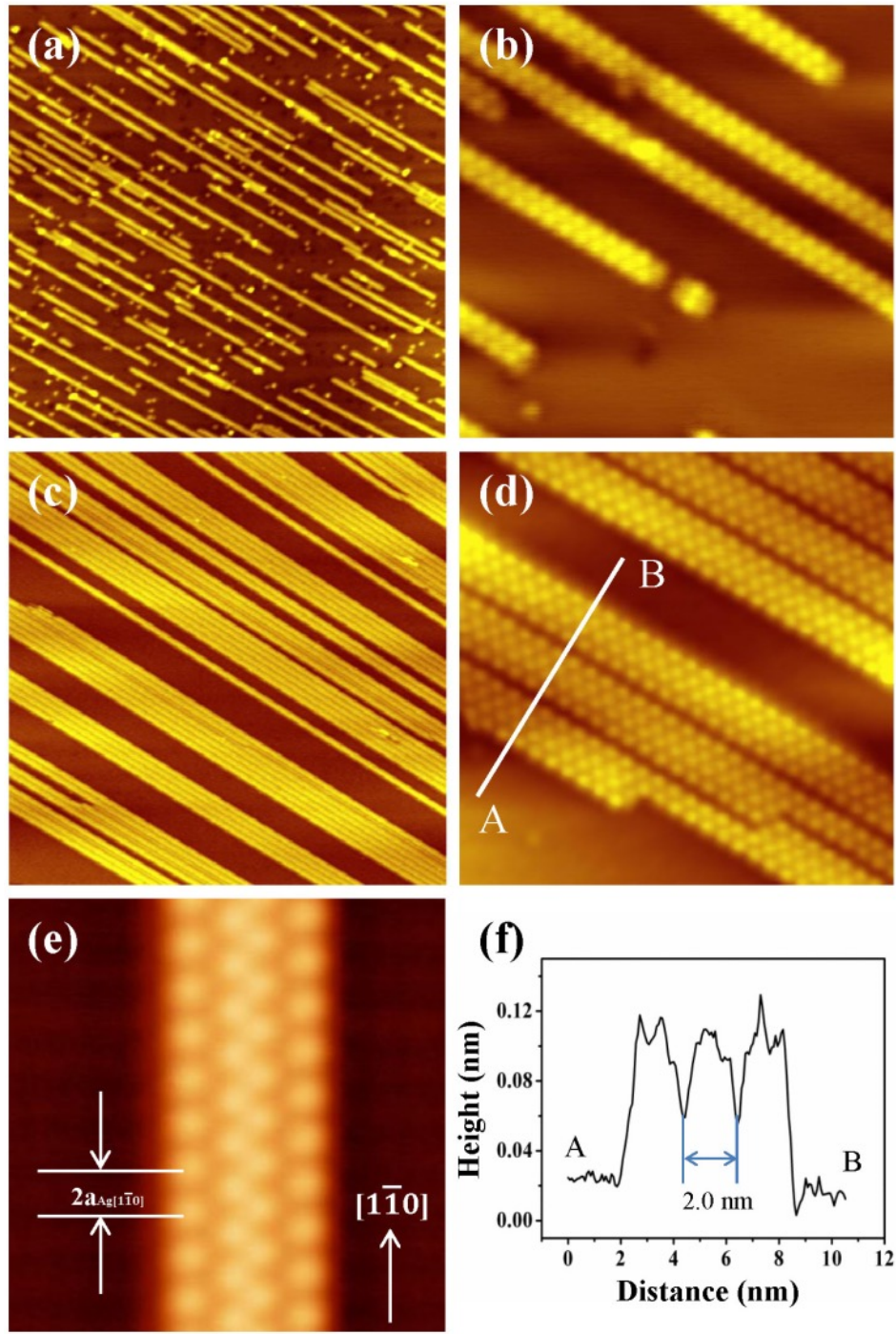

FIG. 1: (color online) (a) 100×100 nm² STM topographic image (V = -1.0 V) of 0.3 ML silicon deposited on Ag(110) at room temperature. (b) High resolution STM topographic image (15×15 nm², V = 1.1 V) of the structure of the 1.0 nm wide SiNRs. (c) 140×140 nm² STM topographic image (V = 1.6 V) of 0.4 ML silicon deposited on Ag(110) at 440 K. (d) High resolution STM topographic image (15×15 nm², V =1.0 V) of the atomic structure of the 2.0 nm wide SiNRs. (e) High resolution STM image showing the atomic structure of Ag(110) and the SiNRs simultaneously. (f) Line profile across the white line in (d) showing the width of the nanoribbon.

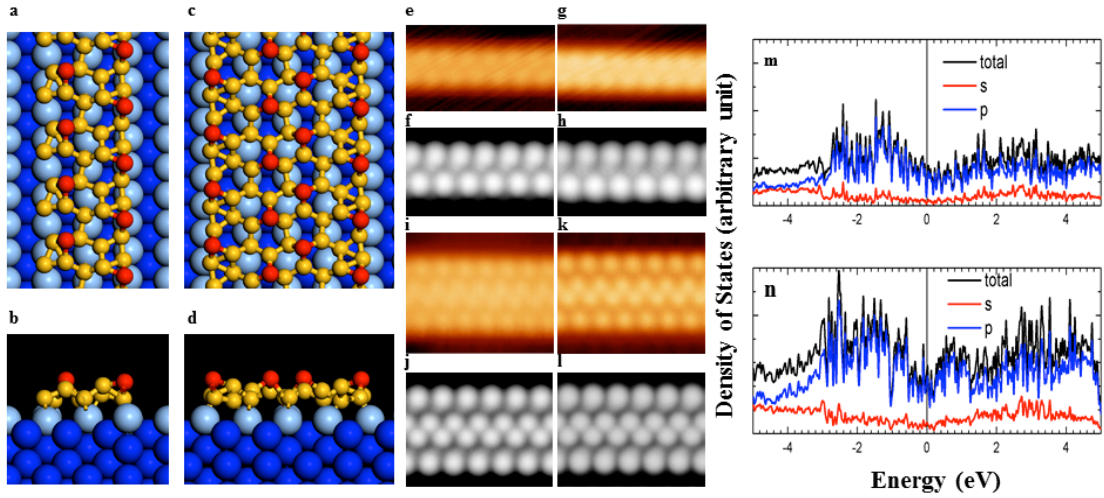

FIG. 2: (color online) (a) and (b) Top and side view of the relaxed structural model of 1.0 nm wide SiNRs on top of Ag(110). (c) and (d) Top and side view of the relaxed structural model of 2.0 nm wide SiNRs on top of Ag(110). Light blue balls: topmost Ag atoms; dark blue balls: underlying Ag atoms; red balls: upper buckled silicon atoms that can be probed by STM; yellow balls: other silicon atoms. (e)-(l) Simulated STM images compared with the experimental STM images at different bias voltages. Bias voltage: (e): 1.1 V; (g): -1.0 V; (i): -1.5V; (k): 1.0 V; (f) and (j): 0-1.5 V; (h) and (l) -1.5-0 V; (m) and (n) Calculated partial density of states of the relaxed SiNRs without Ag(110) for the 1.0 nm and 2.0 nm SiNRs, respectively.

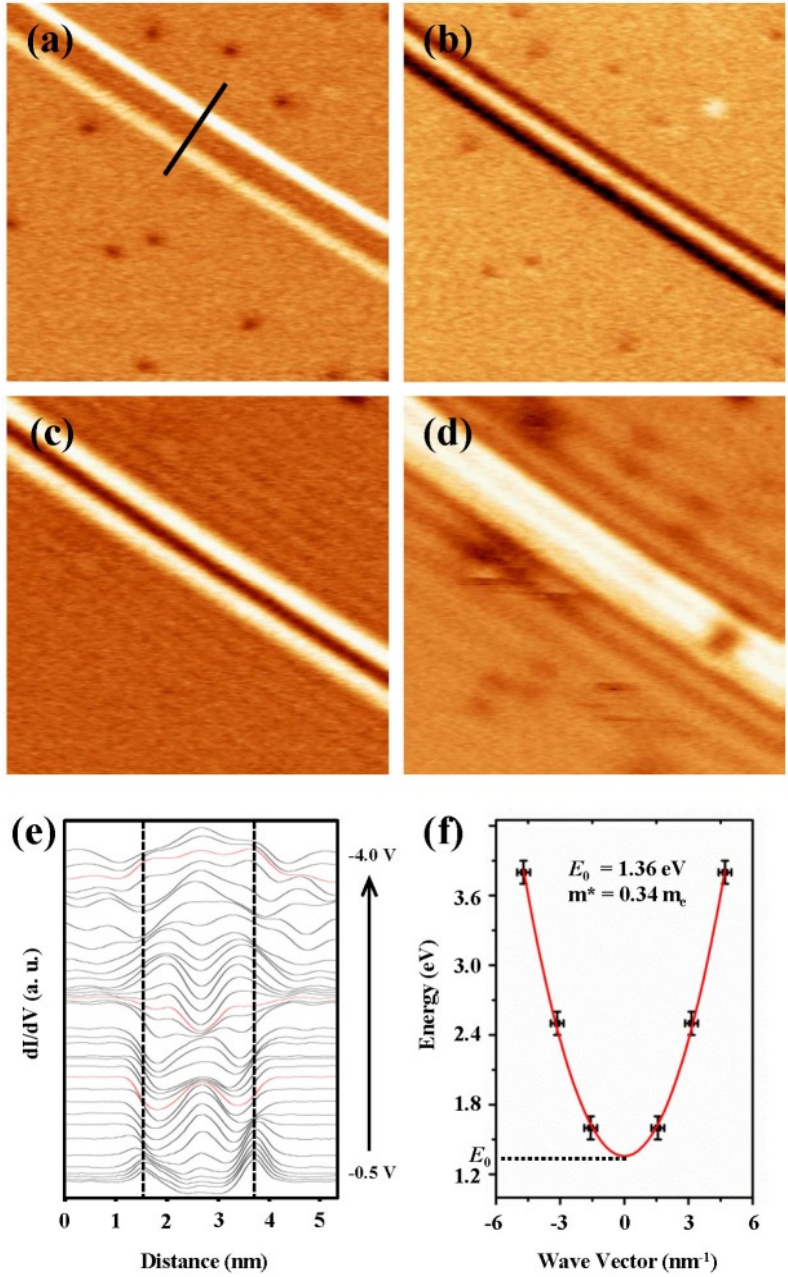

FIG. 3: (color online) dI/dV maps (20×20 nm$^2$) on an isolated 2.0 nm nanoribbon at different bias voltages. (a) -0.8 V. (b) -1.2 V. (c) -2.8 V. (d) -3.8 V. (e) Line profiles perpendicular to the SiNR (along the black line in Fig.2(a)). The vertical coordinates are offset from one another for clarity. The red curves mark pure eigenstates of the quantum well. (f) Energy-momentum dispersion of the electronic states for the 2.0 nm quantum well with each point obtained from the red curve in (e). The red solid line are the best parabolic fit to the data, which yields the onset energy $E_0$=1.36 eV and effective mass $m^*$=0.5m$_e$.